\documentclass[pre,aps,twocolumn,showpacs]{revtex4} 
\usepackage{epsfig} 
\usepackage{bm}

\newcommand{\Onecol} 
{\begin{widetext} 
\onecolumngrid} 
\newcommand{\Twocol} 
{\end{widetext} \twocolumngrid}
 
\newcommand{\be}{\begin{equation}} 
\newcommand{\ba}{\begin{array}} 
\newcommand{\bea}{\begin{eqnarray}} 
\newcommand{\bfi}{\begin{figure}} 
\newcommand{\ee}{\end{equation}} 
\newcommand{\ea}{\end{array}} 
\newcommand{\eea}{\end{eqnarray}} 
\newcommand{\efi}{\end{figure}}

\begin{document}

\title{Stochastic Resonance in Two
DimensionalLandau Ginzburg
Equation}
\author{Roberto Benzi$^1$, Alfonso Sutera$^2$}
\affiliation{$^1$ Dipartimento di Fisica and INFM, Univ. di Roma "Tor Vergata", 
via della Ricerca Scientifica 1, 00173, Roma, Italy.\\
$^2$ Dipartimeto di Fisica Univ. di Roma ``La Sapienza'', P.le Aldo Moro 4, 00185, Roma, Italy.}
\begin{abstract}
We study the mechanism of stochastic resonance in a two
dimensional Landau Ginzburg equation perturbed by a white
noise. We shortly review how to renormalize the equation
in order to avoid ultraviolet divergences. Next we show
that the renormalization amplifies
the effect of the small periodic perturbation in the
system. We finally argue that stochastic
resonance can be used to highlight the effect
of renormalization in spatially extended system
with a bistable equilibria.
\end{abstract}
\maketitle
\bigskip
\bigskip
\bigskip
About twenty years ago, in studying the possible causes of the ice
ages \cite{sutera1}\cite{sutera2}, we introduced the concept of stochastic
resonance \cite{BSV}. This process (hereafter,SR) requires
that a physical system has its fundamental symmetry broken in such a
away that its phase space is not homogeneous. Thereby, the time spent
in different regions of the phase space is not uniformly distributed.
If the system is stochastically forced, the probability of
large fluctuation can be periodically locked to a small
external periodic forcing. The relevant
implications of the SR are the generality of its set up and the
limited band, in the parameter space, wherein the mechanism occurs.
The latter allows for its exploiting in enhancing the signal to noise
ratio in a noisy system, while the former makes the concept
applicables to a large class of systems. Furthermore, we notice that
the mechanism only requires that the system has large deviations properties and,
therefore,
the restriction to systems stochastically perturbed,
is just a case amenable to explicit evaluation
of the resonace range in the parameter space. In practice, a
stochastic forcing is required only for systems of gradient type
(i.e., when a Lyapunov function exists surely) or whenever single
point attractors exist. 

At the time of the publication above
mentioned, we had doubts whether the mathematical
process leading to stochastic resonance were a stochastic process at
all. Fortunately in \cite{f2000} has proved that a wide class of
stochastic process obeying to a SR mechanism exists. Therefore we
feel confident that, at least in a finite dimensional space, the SR
is a well
posed mathematical problem. When we consider stochastically perturbed
infinite dimensional systems the problem may lead to a not consistent
mathematics since
the equations for the backward and the foreward probability density
function may not exist. Moreover, when a white noise in space and
time is considered, the resulting equations are plagued by divergent
correlation functions. In a series of papers \cite{jona1},\cite{jona2},\cite{jona3}, it
was proved that these problems could be overcame by the method of the
renormalisation theory. On the ground of these rigorous, mathematical
results, in the present paper we wish to study the consequences of
applying an external periodic (in time) forcing to a simple infinite
dimensional stochastic equation.

In particular, we discuss the SR phenomenon
in two dimensional Landau Ginzburg equation stochastically
perturbed by a white noise:
\begin{equation}
\label{SR}
\partial_t \phi = \nu \Delta \phi - \frac{\partial V(\phi)}{\partial
\phi} + \epsilon \frac{dW}{dt}
\end{equation}
where
\begin{equation}
V(\phi) = - \frac{1}{2} m \phi^2 - \frac{1}{4} g \phi^4
\label{potential}
\end{equation}
and $\frac{dW}{dt}$ is a white noise delta correletad both in space
and in time.
The field $\phi$ is assumed to be real and defined on the
torus $[0,L]X[0,L]$,$L = 2 \pi$, with periodic boundary conditions.
Notice that the previous equation describes mean field theories
nearby critical points, therefore, the use of white noise in space
and time is mandatory when we do not want to introduce  unknown
physical scales. Moreover, unless $m<0$, the theory has no broken
symmetries.

As we mentioned, the phenomenon of SR has been widely studied in
literature since it was introduced  \cite{BPSV,NIC82}(for a
review see \cite{rev}).
The study of SR in one space dimension, with a broken symmetry, has
been presented in \cite{BSVJPA}. In this case,in fact, the stochastic
equation does not show any particular novelty. As we shall see, the
extension to more spatial dimension leads to interesting
consequences. In particular we will consider their effects on the SR.

As it has been discussed in \cite{BJLS} that equation (\ref{SR}) has
no meaning in
more than one space dimension unless it is renormalized.
In D=2, the renomarlization amounts to change any non linear term in
(\ref{SR}) by its Wick product, a rather well known procedure
in quantum field theory. By using the Wick product,
equations (\ref{SR}) and (\ref{potential}) becomes:
\begin{equation}
\label{SR2}
\partial_t \phi_{\Lambda} = \nu \Delta \phi_{\Lambda} - m
\phi_{\Lambda}
- g \phi_{\Lambda}^3 + 3 g E_m (\Lambda) \phi_{\Lambda} +
\epsilon \frac{dW_{\Lambda}}{dt}
\end{equation}

In eq. (\ref{SR2})
we have introduced the cutoff field $\phi_{\Lambda}$, where
$\Lambda$ is an ultraviolet cutoff in the Fourier space. Namely
by denoting $\Phi (\vec{k})$ the Fourier
transform of $\phi (x) $ we have:
\begin{equation}
\label{cutoff}
\phi_{\Lambda} = \int_{|\vec{k}| < \Lambda} \Phi (\vec{k})
e^{i\vec{k}\vec{x}} d\vec{k}
\end{equation}
and an analogous definition has been used for
$ \frac{dW}{dt}$.
In (\ref{SR2}) the quantity $ E_m (\Lambda) $ is
the expectation value of second order moment of
the field $ Z_{\Lambda}$ satisfying
the linear stochastic differential equation:
\begin{equation}
\partial Z_{\Lambda} = \nu \Delta Z_{\Lambda} - m
Z_{\Lambda} + \epsilon \frac{dW_{\Lambda}}{dt}
\end{equation}
In two space dimensions one has:
\begin{equation}
E_m (\Lambda)  = C \frac{\epsilon^2}{\nu}
ln [ \frac{\nu \Lambda^2 + m}{m} ]
\label{edim}
\end{equation}
where $C$ is a constant order one. Finally, let us recall that the quantity
$ \phi_{\Lambda}^3 - 3 E_m (\Lambda) \phi_{\Lambda}$ is the Wick product of 
$ \phi_{\Lambda}^3$.

One can show \cite{BJLS} that eq. (\ref{SR2}) has a well
defined limit for $\Lambda$ going to infinity.
Let us remark that this limit is achieved by keeping
constant the finite volume. In this case one can show
that equation (\ref{SR}) becomes: \begin{equation}
\label{SRnew}
\partial_t \phi_{\Lambda} =
\nu \Delta \phi_{\Lambda} -
\frac{\partial V_{eff}(\phi_{\Lambda})}{\partial
\phi_{\Lambda}}
+ \epsilon \frac{dW_{\Lambda}}{dt}
\end{equation}
where:
\begin{equation}
V_{eff}(\phi) =  \frac{1}{2} \rho \phi^2 - \frac{1}{4}
g \phi^4 \label{veff}
\end{equation}
The value of $\rho$ is determined by the following set
of equations: \begin{equation}
\rho = - m + 3g [ E_m(\Lambda) - E_{\alpha}(\Lambda)]
\label{rho}
\end{equation}
\begin{equation}
\alpha = \frac{\partial^2 V_{eff}}{\partial \phi^2}
\quad at \quad \phi=\phi_e
\label{alpha}
\end{equation}
where $\phi_e$ is the equilibria of the effective
potential, i.e. $\phi_e$ is
determined by solving the equation:
\begin{equation}
\frac{\partial V_{eff}}{\partial \phi} = 0
\label{solstaz}
\end{equation}
The set of equations
(\ref{rho},\ref{alpha},\ref{solstaz}) have a clear
physical
meaning \cite{BJLS}, shortly reviewed as follows.
First of all, let us note that $\rho$ is independent of
$\Lambda$ in the limit $\Lambda \rightarrow
\infty$ because of eq. (\ref{edim}). Eq.
(\ref{rho}) is a redefinition of the Wick product
in terms of a new constant $\alpha$, which is
defined in
equation
(\ref{alpha})
as
the curvature of the effective potential at one
of its equilibrium values, namely $\phi_e$.  The
physical idea of eq. (\ref{alpha})  is that, the
counter term
introduced in order to renormalize the Landau
Ginzburg equation, produces a double well
potential. The position of the new minimum,
determined by $\rho$, must be consistent with the
fluctuations around the minimum itself.

Using equations (\ref{rho},\ref{alpha},\ref{solstaz}) one gets
a nonlinear equation for the variable $x = \frac{\alpha}{m}$,
namely \begin{equation}
x = - 2 + 2 K log x
\label{logx}
\end{equation}
where $ K = \frac{3gC\epsilon^2}{m\nu}$. Let us remark, for
future purpose, that it is possible to show that
a solution of eq. (\ref{logx}) always satisfies $x > 2 K $.
As discussed in \cite{BJLS}, the effective potential describes
rather well the statistical properties of the space averge
field of eq.(\ref{SR}).

In general we can consider the variables $\Phi (l) $ defined
as the average  in space of $\phi(x)$ on a box of side $l$.
For $l < L = 2 \pi $, the statistical properties of $\Phi (l)$
are no longer described by the same value of $\rho$
because of the increasing effect, at small scales, of
the stochastic fluctuations. In particular,
for $l \rightarrow 0$, one finds that the statistical
properties of $\Phi (l) $ are described by a single
well potential centered around $\Phi (l) = 0$.

We want now to discuss the solutions of eq. (\ref{SR})
when a small periodic perturbation is added to the
system and in particular
we want to discuss the effect of renormalization of the
mechanism on the stochastic resonance.

To this aim, following the original discussion given in
\cite{BSV} we consider the case of a time independent
constant $A$ added to the r.h.s. of (\ref{SR}). The
crucial point is the computation of the
effective potential $V_{eff}$ previously introduced.
Equation (\ref{rho}) are unchanged while equations
(\ref{alpha},\ref{solstaz}) become:
\begin{equation}
\alpha = - \rho + 3 g {\phi_e}^2
\label{newalpha}
\end{equation}
\begin{equation}
\rho \phi_e - g {\phi_e}^3 + A = 0
\label{phie}
\end{equation}
The effect of the constant perturbation is to change the
equilibrium solution of the effective potential and therefore
to change the fluctuations around the equilibrium solution,
which in
turn change the value of  $\rho$. The latter effect is due
only
to renormalization.

In order to understand qualitatively the contribution due to
renormalimation,
one can compute a perturbative solution of the set of
equations (\ref{rho},\ref{newalpha},\ref{phie}) in power of
$A$. At first order one obtains:
\begin{equation}
\rho_{1,2} = \rho_0 + \frac{6 g \phi_{e0}H}{1-2H}
\frac{A}{2 \rho_0} \label{rho12}
\end{equation}
\begin{equation}
\phi_{e1,e2}= \phi_{e0} + \frac{1+H}{1-2H} \frac{A}{2
\rho_0} \label{phi12}
\end{equation}
where $H= \frac{3gC\epsilon^2}{\nu \alpha_0}$
( $ \alpha_0 = 2 \rho_0 $)
and all quantities with the index $0$ refer to values
computed for $A=0$. The previous definition of $H$ implies
$H = \frac{K}{x} < \frac{1}{2}$ because of the discussion
on the solution of eq. (\ref{logx}).
Let us remark that we obtain two different values of $\rho$
and $\phi_e$
because two different values of $\phi_{e0}$ are possible,
one negative and the other one positive. Thus there exists
two possible effective potential, each one describing the
two well respectively.

By using equations (\ref{rho12}) and (\ref{phi12}) is now
possible to compute the effective potential difference
between the two stable solutions and the unstable solution.
One gets: 
\begin{eqnarray}
\Delta V_{1,2} = -\frac{\rho^2}{4g} + A_R \phi_{e0} \label{master} \\
A_R = A (1 + \frac{3H}{2(1-2H)}) \label{AR}
\end{eqnarray}
In the limit of $\epsilon \rightarrow 0 $ we have $H
\rightarrow 0$ and the effect of renormalization disappears.
For finite value
of the noise and for $g$ large enough the renormalization
amplifies the asymmetry of the double well potential.
Therefore, we should expect that, in two dimensions
Landau Ginzburg equation, the mechanism of SR acts for smaller
value of the forcing amplitude with respect to the
one estimated by looking at the equilibrium
probability distribution with no periodic forcing.

In order to test the renormalization effect on SR, we
have performed a numerical simulation on square
lattice of 16x16 points.
The parameters of our numerical simulation are
$g = 220$, $\nu = 0.1 $, $m = 0.1 $, $\epsilon =
0.1$. On the lattice, the quantity $E_m(\Lambda)$, defined in eq. (\ref{edim}), is given by:
\begin{equation}
E_m(a) = \Sigma_{p,q} \frac{a^2\epsilon^2}{2S(m,\nu,q,p)}
\label{edia}
\end{equation}
where
$$
S(m,\nu,q,p) = ma^2 +4\nu-2\nu cos(2\pi q/N) -2\nu cos(2\pi p/N)
$$
In equation (\ref{edia} we introduce the lattice mesh 
$a= 2\pi / N$. It follows that  $\Lambda = a^{-1}$.

For different values of $A$, by applying Newton method,
we have computed  the solution of  equations
(\ref{rho},\ref{newalpha},\ref{phie}), using (\ref{edia}).
It turns out that, with our numerical parameters,
$\rho_0 = 4.9.$ and $A_R \sim 2 A$, i.e, the external forcing is amplified by
almost  a factor 2.

\begin{figure} 
\centering 
\includegraphics[width=.5\textwidth]{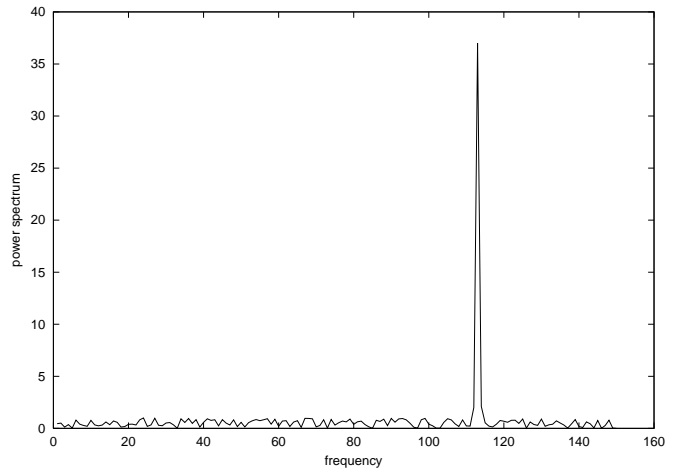} 
\caption{Power spectrum for the space average field  $\Phi (L)$
obtained for the numerical simulation of eq. (\ref{SR}). The
peak corresponds to a period of 18 time units.} 
\label{fig1a} 
\end{figure}

We have performed a long time integration of
(\ref{SR}) with $A=0$. The equilibrium probability
distribution of $\Phi (L)$ has been found to be
bimodal with maxima centered
at $ \pm \sqrt{\frac{\rho_0}{g}} $
and the average transition time $\tau$ between the
two equilibrium $\tau = 9$ in time units of $m^{-1}$.
Next, using the analytical theory of SR discussed in
\cite{BSV},
we have computed,
by taking into account
the amplification of the renormalization,
the value of $A$
for which the system should show a clear SR. We have
found $A = 0.1$. Finally we have applied a periodic
perturbation on the r.h.s. of eq. (\ref{SR}) with
amplitude $A=0.1$ and period equal to
$2\tau= 18$, i.e. two times the average transition time.
We have performed a long time integration corresponding
to 2048 time units. In figure (\ref{fig1a}) we show the
power spectrum of the average field $\Phi (L)$. As one
can clearly seen, there is an extremely well defined
peak at frequency 113 which corresponds to a period of
18 time units.

\begin{figure} 
\centering 
\includegraphics[width=.5\textwidth]{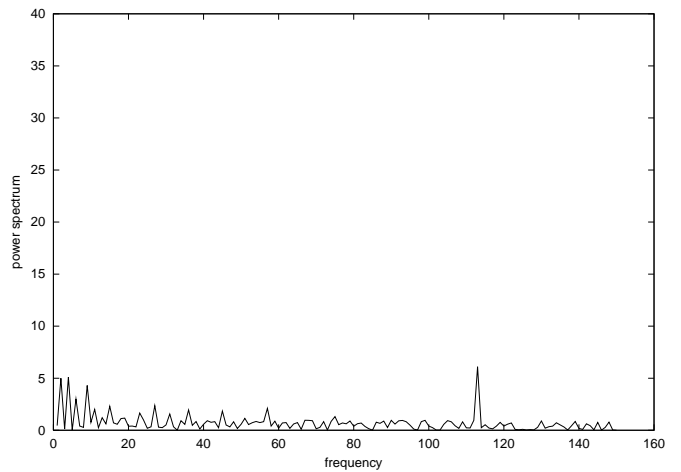} 
\caption{Same as in fig. 1b for wero dimensional stochastic
differential equation (\ref{xD}). Note that the amplitude of
the peak at period of 18 time units is much smaller with
respect to figure (\ref{fig1a}) } 
\label{fig1b} 
\end{figure}

In order to show the amplification effect on SR due to
renormalization,
we have numerically integrated the zero dimensional
stochastic differential equation:
\begin{equation}
\frac{dx}{dt} = \rho_0 x - g x^3 + A
cos(\frac{2t\pi}{18}) + \sigma \frac{dW}{dt}
\label{xD}
\end{equation}
where $A=0.1$ as before and the variance $\sigma$ of the
white noise has been tuned in order to reproduce an
average transition time of $9$ ($\sigma = 0.0165$) for
$A=0$. With this choice of $\sigma$ we have been able to
reproduce the equilibrium probability distribuiont of
$\Phi (L)$ at $A=0$ for the 2D Landau Ginzburg equation.
As for the previous case, we have integrated
eq. (\ref{xD}) for $2048$ time units. In figure (\ref{fig1b}) we show
the power spectrum of $x$. At variance with figure (\ref{fig1a}),
only a small
effect of the periodic forcing is felt by the system, i.e.
the system does not exhibits any SR.

The comparison between figures (\ref{fig1a}) and (\ref{fig1b}),
indicates clearly that the renormalization effect is
acting as an amplifier of the external forcing, as
previously discussed in the framework of a perturbation
theory.

It is quite interesting to compute the statistics of the
transition times between the minima at different scales as
described by the variables $\Phi (l)$. In absence of
forcing the probability distribution
of the transition time for
$\Phi (L)$ is exponential \cite{BJLS}. In the stochastic
resonance the probability distribution of the transition
time for $\Phi (L)$ should show a well defined maximum at 9
time units.
On the other hand, if we consider the transition time for
$\Phi (l)$ with $l \ll L $, then we expect that the forcing
is not
able
to produce any SR. This is due to the fact that the
parameter describing the effective potential for
small $l$ are changed and, in particular, the
average transition time between
the minima becomes much smaller \cite{BJLS}.
This effect is indeed observed in our numerical
simulation. In fig. (2) we show the probability
distribution
of the transition time for $\Phi (L)$ and $\Phi
(\frac{L}{16})$. As predicted, for the small scale, the probability
distribution of the transition time is exponential.
\begin{figure} 
\centering 
\includegraphics[width=.5\textwidth]{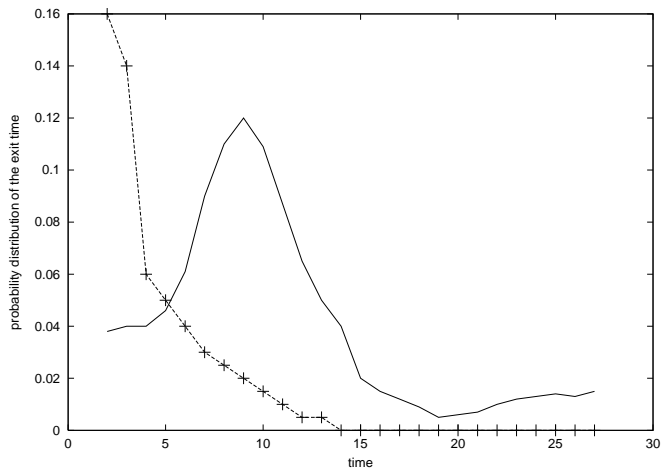} 
\caption{Probability distribution of the transition time for $\Phi
(L)$, continuos curve, and $\Phi (\frac{L}{16}) $,
continuos curve with symbols.} 
\label{fig2} 
\end{figure} 
We can therefore reach the conclusion that, becuase of the renormalization, the SR mechanism
becomes scale depedent, a feature not previously observed.

In this letter
we have shown that renormalization can amplify the SR
in two
dimensional Landau Ginzburg equation. We want to
remark that this effect can also be used to test, in a
given experimental situation, whether or not a
renormalization mechanism is acting in the system. In
this case SR can be used as a tool
to measure quantitatively the effect, if any, of
renormalization in spatially exnteded systems. We
remark that the present effect may be not restricted
to system of gradient type and, in principle, could be detected in
spatially extended deterministic systems.

\bigskip
{\bf Acknowledgements}
We thank Prof. Jona Lasinio for useful
discussions.


\end{document}